\documentclass[11pt]{article}

\usepackage[numbers,sort&compress]{natbib}	%Sortiert;Zitate;und;macht;aus;[10,11,12,13,14];[10-14]
\usepackage{hypernat}				%Macht;hyperref;und;natbib;kompatibel
\usepackage{graphicx}		
\usepackage{amssymb}
\usepackage{simplemargins}
\setallmargins{1in}
\usepackage[american]{babel}
\usepackage{amsmath}
\usepackage{amsfonts}	
\usepackage{amsmath}
\usepackage{url}
\usepackage{multicol}
\usepackage[table]{xcolor} \definecolor{lightyellow}{HTML}{FFFFBB}
\usepackage[pdftex]{hyperref}
\hypersetup{
  unicode=false,
  pdffitwindow=true,
  pdfpagelayout=TwoPageLeft,
  pdfhighlight=/P,
  pdfmenubar=false,
  pdftoolbar=false,
  pdfstartview=FitV,
  pdfnewwindow=true,
  colorlinks=true,
  linkcolor=blue,
  anchorcolor=black,
  citecolor=blue,
  filecolor=black,
  menucolor=black,
  urlcolor=blue,
  breaklinks=true,
  pdftitle={Some Indications on how to Calibrate the Social Force Model of Pedestrian Dynamics},
  pdfkeywords={Pedestrian, Crowd, Mass, Simulation, Dynamics, Viswalk, Social Force Model, Calibration}
  pdfsubject={Social Force Model, Pedestrian, Crowd, Mass, Simulation, Dynamics, Viswalk, Vissim, Calibration, Model Simplification},
  pdfauthor={Kretz},
  pdfcreator={Tobias Kretz},
  pdfproducer={T. Kretz},
}

\title{Some Indications on how to Calibrate the Social Force Model of Pedestrian Dynamics}

\author{Tobias Kretz, Jochen Lohmiller, Peter Sukennik\\
PTV Group, Haid-und-Neu-Stra{\ss}e 15, D-76131 Karlsruhe, Germany\\
\texttt{\{First.Family\}@ptvgroup.com}\\
}%

\begin{document}

\maketitle

\abstract{
The Social Force Model of pedestrian dynamics is formulated in a way that a) most of its parameters do not have an immediate interpretation, b) often one single parameter has an impact on many aspects of walking behavior and c) a certain aspect of walking behavior results from the values of more than one parameter. This makes calibration difficult. The aim of this paper is to give practitioners an indication of how to proceed in the calibration process. For this by analytical transformations the parameters of the Social Force Model are related to real properties that have a clear and immediate meaning and which are also highly relevant result properties of a simulation: extent and clearance time of a queue, respectively maximum density and capacity flow. The theory for this is presented as well. As a side effect it can give a deeper understanding of the model for everyone interested in theoretical aspects.
}

\section{Introduction and Motivation}
The Social Force Model (SFM) \cite{helbing1995social,Helbing2000simulating,johansson2007specification,helbing2011pedestrian} is one of the most prominent examples of a model of pedestrian dynamics. It is formulated in a way that a) most of its parameters do not have an immediate interpretation in the sense that they cannot be measured directly, b) often one single parameter has an impact on many aspects of walking behavior and c) a certain aspect of walking behavior results from the values of more than one parameter. In this regard the SFM is very different from a number of car-following models like the Wiedemann model \cite{wiedemann1974simulation} where nearly each parameter has an immediate interpretation (can be measured directly) and has only a local effect; ``local'' in the sense that within the model it affects only a certain observable properties of driving behavior, i.e. of the results. 

This structure of the SFM makes calibration challenging, since when the meaning of a parameter is abstract instead of immediate, for someone confronted with the model for the first time, it is often unclear which are the most relevant parameters to adjust and even in which direction a certain modification of the value of a certain parameter changes the simulation results. Among models of pedestrian dynamics the SFM is not unique in having parameters with a rather abstract meaning. The cellular automata floor field models for example share that property to a similar degree \cite{burstedde2001simulation,nishinari2004extended,kretz2008counterflow}.

Compared to action-point car-following models the SFM also has an advantage: it can be analyzed, transformed, and simplified in a rigid mathematical way allowing a deeper understanding of the a priori abstract parameters, which leads to the aim of this paper: to give practitioners an indication of how to proceed in the calibration process. For this by analytical transformations the parameters of the SFM are related to real properties that have a clear and immediate meaning: number of pedestrians, stand still density, and capacity flow. The latter two are highly relevant since they frame the fundamental diagram and thus are important properties in a calibration process. 

At the same time these are basic properties such that a model which reproduces these well does not necessarily work well in more complex situations like  bi-directional flow situations with emergent lane formation. Thus the work done in this contribution is necessary for calibration, but not complete. However, with regard to the numerous variants of the SFM \cite{yu2005centrifugal,lakoba2005modifications,johansson2007specification,yu2007modeling,parisi2009modification,steffen2010modification,zanlungo2011social,ratsamee2012modified,kretz2015social} and to general force-based models \cite{chraibi2010generalized,campanella2010improving} the basic character of the proposed method is beneficial, since it means that it applies to most of these variants.

The outline is as follows: in the next section we will derive the results step by step and also document the model simplifications that lead to the results. A reader not interested in the derivation may skip the section and continue in the ``Summary of Results and Discussion'' section, after which follows a section with examples.

\section{Derivation of Results}
Our starting point is the circular specification of the SFM. It is the simplest version since inter-pedestrian forces depend only on distance between pedestrians. Since in all states of the system that we use in our reasoning relative velocities between pedestrians are zero, the arguments also hold for the elliptical specification II which in addition to distance also considers relative velocity as determining property for the forces. Elliptical specification I, on the contrary, which takes into account distance and the velocity of the pedestrian who exerts the force, is fundamentally different \cite{johansson2007specification}.

The circular specification of the SFM -- neglecting forces from walls -- is defined as:
\begin{equation}
\ddot{\vec{x}}_i(t)=\frac{\vec{v}_{0,i}-\dot{\vec{x}}_i(t)}{\tau_i}+\tilde{A}_i \sum_{j} w(\vec{x}_i(t),\vec{x}_j(t),\dot{\vec{x}}_i(t),\lambda_i) e^{-\frac{|\vec{x}_j(t)-\vec{x}_i(t)|-R_i-R_j}{B_i}}\hat{e}_{ij} \label{eq:circularfull}
\end{equation}
with
\begin{equation}
w(\vec{x}_i(t),\vec{x}_j(t),\dot{\vec{x}}_i(t),\lambda_i)=\lambda_i + (1-\lambda_i) \frac{1+\cos(\theta_{ij}(\vec{x}_i(t),\vec{x}_j(t),\dot{\vec{x}}_i(t)))}{2}
\end{equation}
where $v_{0,i}$ is the desired speed of pedestrian $i$. $\tilde{A}_i>0$, $B_i>0$, $0\leq\lambda_i\leq 1$, and $\tau_i>0$ are parameters of the model. $R_i$ denotes the body radius of a pedestrian. $\hat{e}_{ij}$ has the direction from pedestrian $j$ on pedestrian $i$. $\vec{x}_i$ is the position of a pedestrian and dots mark time derivatives (i.e. speed and acceleration). The sum runs over all -- potentially infinitely many -- pedestrians in a simulation scenario. Function $w()$ is there to suppress forces acting from behind. Within it $\theta_{ij}$ is the angle between pedestrian $i$'s velocity vector and the line connecting pedestrians $i$ and $j$.

Abstract parameters in the sense of the introduction are particularly $A_i$, $B_i$, and $\lambda_i$ and -- to a lesser degree -- $\tau_i$. The desired speed $v_{0,i}$ on the contrary in this contribution and in accordance with most publications on the SFM is viewed as a parameter that has an immediate interpretation and can be measured directly by measuring the free, unobstructed walking speed, although it is an unproven hypothesis that for pedestrians there is a situation independent desired speed that matches the free walking speed. Therefore in the framework of this contribution we assume that there is no freedom in choosing the value of $v_{0,i}$ and count it to the empirical properties although at the same time it is a model parameter.

From here on we assume that parameters $\tilde{A}$, $B$, $\lambda$, $\tau$ $R$, and $v_0$ have identical value for all pedestrians, so we omit the indices. This allows to combine $\tilde{A}$ and $R$ into a new parameter:
\begin{equation}
A=\tilde{A}e^{\frac{2R}{B}}\label{eq:AAtilde}
\end{equation}

It is essential to keep this parameter transformation in mind, since some implementations of the SFM require $\tilde{A}$ and some $A$ as input parameter. Put differently: if there is no $2R$ in the exponent and if ``distance between pedestrians'' refers to surface to surface one needs to calculate with parameter $\tilde{A}$, while if it is center to center parameter $A$ is required. In this contribution ``distance'' always means center to center. We use the formulation with $A$ for theoretical investigation, but for all simulations $A$s mean $\tilde{A}$, i.e. the $2R$ are ``added'' in the simulation.

Since it is obvious which properties are time dependent, we also omit the ``$(t)$''. Then equation (\ref{eq:circularfull}) can be written for the one-dimensional case:
\begin{eqnarray}
\ddot{x}_i         &=& \frac{v_0-\dot{x}_i}{\tau}+A \sum_j w(x_i,x_j,\lambda) e^{-\frac{d_{ij}}{B}} \label{eq:circular1d}\\
d_{ij}             &=& |x_j-x_i| \\
w(x_i,x_j,\lambda) &=& \lambda \text{ if } x_j-x_i<0\\
w(x_i,x_j,\lambda) &=& -1 \text{ if } x_j-x_i>0
\end{eqnarray}
with the additional assumption that for all pedestrians and times $v_0, \dot{x}>0$. This can be done without limiting generality since we are not interested in phenomenons far from equilibrium, but only investigate a stationary state of the system (strong equilibrium).

Considering only nearest neighbor interactions (one neighbor $i+1$ ahead and one neighbor $i-1$ behind) and neglecting all other forces the sum in equation (\ref{eq:circular1d}) is reduced:
\begin{equation}
\ddot{x}_i = \frac{v_0-\dot{x}_i}{\tau}- A \left(e^{-\frac{d_{i,i+1}}{B}} - \lambda e^{-\frac{d_{i,i-1}}{B}}\right) \label{eq:circular1dNN}\\
\end{equation}
for all pedestrians who have a leading and a following pedestrians. For the first and the last pedestrians in the line one of the two terms in brackets is missing. 

Assume now the first pedestrian in line is waiting in front of a red traffic signal. Let us define its position being $x_0=0$. For a finite system (i.e. there is a last pedestrian) and if $\lambda>0$ the distances between pedestrians grow towards the end of the queue since the last pedestrian is not experiencing a pushing force from behind, therefore exerting a smaller force to his leader who therefore is further away from his leader than others further down in the queue. The effect is most pronounced toward the end of the queue, while distances vary only slightly toward the front, particularly if $\lambda$ has a value close to zero, much smaller than 1. We assume now that there is a long queue of $Z$ pedestrians and we are interested in the distance $L$ from the first to the $N$th pedestrian where $N<<Z$ when all pedestrians are standing still and for each pedestrians there is zero net force, i.e. the left hand side of equation (\ref{eq:circular1dNN}) is zero. Because $N<<Z$ for pedestrians in concern we only make a small mistake assuming -- as if the queue was infinitely long -- that all distances between pedestrians are identical setting $d=d_{i,i+1}=d_{i,i-1}$:

\begin{equation}
0 = \frac{v_0-\dot{x}_i}{\tau}- (1 - \lambda) A e^{-\frac{d}{B}}  \label{eq:circular1dNNapprox}\\
\end{equation}

where the left side (acceleration) is zero, since we are at equilibrium. This can be solved easily for $d$; and as stated we are interested in the smallest possible distance $d_{min}$, i.e. when speed is zero:

\begin{eqnarray}
d_{min} &=& B \ln(\alpha) \\
\alpha  &=& \frac{(1-\lambda) A \tau}{v_0} \label{eq:alpha}
\end{eqnarray}
with parameter combination $\alpha$ defined for convenience since this parameter combination will occur often in the remainder.

The inverse of $d_{min}$ is the maximum density $\rho_{max}$ which we identify with the stand still density, i.e. the upper end point in the fundamental diagram. For this and for above mentioned $L$ results trivially:
\begin{eqnarray}
\rho_{max} &=& \frac{1}{B \ln(\alpha)} \label{eq:rhomax} \\
L &=& (N-1) B \ln(\alpha)              \label{eq:L}
\end{eqnarray}

Equation (\ref{eq:L}) connects the two properties $L$ and $N$ which have an immediate-non-abstract meaning with the parameter $B$ and parameter combination $\alpha$ of the SFM and give a first indication which parameter modification has which impact on simulation results.

The second step is to imagine that the traffic light which before stopped pedestrians turns green and that the queue of pedestrians discharges. At first pedestrians which cross $x=0$, i.e. the stop line of the signal, are still accelerating, yet after some time, the fourth, or fifth, or tenth pedestrian has already accelerated to the speed which is possible at that density. The further discharge occurs with capacity flow $j_c$ at capacity density $\rho_c$.

Also for this situation we assume that distances between pedestrians are constant. We resolve equation (\ref{eq:circular1dNNapprox}) for $\dot{x}_i$ and use it for $v$ in the fundamental equation:

\begin{eqnarray}
j(\rho)&=&\rho v(\rho) \label{eq:fundamental} \\
       &=&\rho v_0 \left(1 - \alpha e^{-\frac{1}{B\rho}}\right) \label{eq:fundamentalv}
\end{eqnarray}

At capacity ($\rho_c$, $j_c$) the first derivative of $j$ by $\rho$ vanishes:
\begin{eqnarray}
\frac{\delta j}{\delta \rho} &=& v_0 \left(1 - \alpha \left(1 + \frac{1}{B\rho}\right) e^{-\frac{1}{B\rho}}\right) = 0 \text{, for } \rho=\rho_c
\end{eqnarray}

\newcommand{\W}{\operatorname{W}}

Next we multiply and at the same time divide the last term by $e$:
\begin{equation}
0 = v_0 \left(1 - \alpha e \left(1 + \frac{1}{B\rho_c}\right) e^{-\left(1 + \frac{1}{B\rho_c}\right)}\right)
\end{equation}
which prepares the equation to be resolved for $\rho_c$:
\begin{equation}
\rho_c = -\frac{1}{B \left(1 + \W\left(-\frac{1}{\alpha e}\right)\right)} \label{eq:capacitydensity}
\end{equation}
where $\W()$ is  the Lambert W function \cite{corless1997sequence,barry2000analytical,chapeau2002numerical,scott2014asymptotic,WeissteinLambertW} which is defined as inverse of $x e^x$, i.e. 

\begin{equation}
x = \W(x)e^{\W(x)}. \label{eq:LambertW}
\end{equation}

To be precise: in this work $\W()$ always references the lower real valued branch of the Lambert W function which is denoted as $W_{-1}()$ when there is a need to distinguish different branches.

The lower branch of $\W()$ is defined for all $-1/e \leq x < 0$. This is the case for the argument $-1/e/\alpha$, since equation (\ref{eq:rhomax}) requires $\alpha > 1$. All of its values are $\W()<-1$ which is what is also required in equation (\ref{eq:capacitydensity}).

Using this in equation (\ref{eq:fundamentalv}) we get as capacity flow:
\begin{eqnarray}
j_c &=& -\frac{v_0}{B \left(1 + \W\left(-\frac{1}{\alpha e}\right)\right)} \left(1 - \alpha e^{1 + \W\left(-\frac{1}{\alpha e}\right)}\right) \\
    &=& -\frac{v_0}{B \left(1 + \W\left(-\frac{1}{\alpha e}\right)\right)} \left(1 - (\alpha e) \frac{\W\left(-\frac{1}{\alpha e}\right)}{\W\left(-\frac{1}{\alpha e}\right)}e^{\W\left(-\frac{1}{\alpha e}\right)}\right) \\
		&=& -\frac{v_0}{B \left(1 + \W\left(-\frac{1}{\alpha e}\right)\right)} \left(1 - (\alpha e) \frac{-\frac{1}{\alpha e}}{\W\left(-\frac{1}{\alpha e}\right)}\right) \\
		&=& -\frac{v_0}{B \left(1 + \W\left(-\frac{1}{\alpha e}\right)\right)} \left(1 + \frac{1}{\W\left(-\frac{1}{\alpha e}\right)}\right) \\
    &=& -\frac{v_0}{B} \frac{1}{\W\left(-\frac{1}{\alpha e}\right)} \label{eq:capacityflow}
\end{eqnarray}
where for the second transformation we use the defining equation (\ref{eq:LambertW}) of the Lambert W function.

To put it in terms of directly observable properties: the time $T$ for $N$ consecutive pedestrians to cross the stop line with this results trivially in
\begin{eqnarray}
T &=& \frac{(N-1)}{j_c}\\
  &=& - (N-1) \frac{B}{v_0} \W\left(-\frac{1}{\alpha e}\right)
\end{eqnarray}

Equations (\ref{eq:rhomax}) and (\ref{eq:capacityflow}) or their repackings in terms of $L$ and $T$ can be combined to get an equation without parameter $B$ or without parameter $\alpha$. For this we first define a combination of observable properties:
\begin{equation}
q := \frac{j_c}{v_0\rho_{max}} = \frac{L}{v_0T}
\end{equation}

Then $q$ can be written depending only on $\alpha$, but not $B$:
\begin{equation}
q = - \frac{\ln(\alpha)}{\W\left(-\frac{1}{\alpha e}\right)} \label{eq:noB}
\end{equation}
or on $B$, but not $\alpha$:

\begin{equation}
\frac{1}{q\rho_{max}} = \frac{T v_0}{N-1} = -B \, W\!\left(-e^{-\left(1+\frac{1}{B\rho_{max}}\right)}\right) \label{eq:noalpha}
\end{equation}

The right hand side of equation (\ref{eq:noB}) has a lower limit of 0 and an upper limit of 1 since $\W()$ can be approximated \cite{veberic2010having} by
\begin{equation}
\W(x) \approx \ln(-x) - \ln(-\ln(-x)) \text{ for } x \rightarrow 0^-
\end{equation}
and since

\begin{equation}
\lim_{\alpha \rightarrow \infty} \left[- \frac{\ln(\alpha)}{\ln\left(\frac{1}{\alpha e}\right)-\ln\left(-\ln\left(\frac{1}{\alpha e}\right)\right)}\right] = 1
\end{equation}

The lower limit of 0 does not have any relevant implication, yet the upper limit of 1 implies that under the assumptions and simplifications made in this contribution the SFM can only reproduce pedestrian system dynamics where:
\begin{equation}
q<1.
\end{equation}

It appears that this is not a relevant limitation since in \cite{seyfried2005fundamental} roughly the values $v_0\approx 1.25$ m/s, $j_c\approx 0.8$ s$^{-1}$, and $\rho_{max}\approx 2.0$ m$^{-1}$ were found such that $q\approx 0.32$.

Equations (\ref{eq:noB}) and (\ref{eq:noalpha}) can be solved for $\alpha$ resp. $B$. Both transformations include steps which are not obviously goal directed. Therefore we will do this step by step, beginning with equation (\ref{eq:noB}):

\begin{eqnarray}
                                                                                        q &=& - \frac{\ln(\alpha)}{\W\left(-\frac{1}{\alpha e}\right)} \\
                                                        \W\left(-\frac{1}{\alpha e}\right) &=& -\frac{\ln(\alpha)}{q} \\
                                                                      -\frac{1}{\alpha e} &=& -\frac{\ln(\alpha)}{q}e^{-\frac{\ln(\alpha)}{q}} \\
                                                                              \frac{1}{e} &=& \frac{\ln(\alpha)}{q}e^{-\frac{\ln(\alpha)}{q}+\ln(\alpha)}  \\
                                                                            \frac{q-1}{e} &=& \frac{q-1}{q}\ln(\alpha)e^{\frac{q-1}{q}\ln(\alpha)}  \\
                                               \frac{q-1}{e}e^{-\frac{q-1}{q}\ln(\alpha)} &=& \frac{q-1}{q}\ln(\alpha)  \\
                                           e^{\frac{q-1}{e}e^{-\frac{q-1}{q}\ln(\alpha)}} &=& e^{\frac{q-1}{q}\ln(\alpha)}  \\
\frac{q-1}{e} e^{-\frac{q-1}{q}\ln(\alpha)}e^{\frac{q-1}{e}e^{-\frac{q-1}{q}\ln(\alpha)}} &=& \frac{q-1}{e} \\
                                               \frac{q-1}{e}e^{-\frac{q-1}{q}\ln(\alpha)} &=& \W\left(\frac{q-1}{e}\right) \\
                                                     \frac{q-1}{e}\alpha^{-\frac{q-1}{q}} &=& \W\left(\frac{q-1}{e}\right) \\
                                                                                  \alpha  &=& \left[-\W\left(-\frac{1-q}{e}\right) \frac{e}{1-q}\right]^{\frac{q}{1-q}}
\end{eqnarray}

Now for equation (\ref{eq:noalpha}). It can be written as
\begin{eqnarray}
\frac{1}{qB\rho_{max}} &=& - \W\left(-e^{-\left(1+\frac{1}{B\rho_{max}}\right)}\right)\\
-\frac{1}{qB\rho_{max}} e^{-\frac{1}{qB\rho_{max}}} &=& -e^{-\left(1+\frac{1}{B\rho_{max}}\right)}\\
\frac{1}{B}e^{\frac{1}{B\rho_{max}}-\frac{1}{qB\rho_{max}}} &=& \frac{q\rho_{max}}{e}\\
\frac{1}{B}\left(\frac{1}{\rho_{max}}-\frac{1}{q\rho_{max}}\right)e^{\frac{1}{B\rho_{max}}-\frac{1}{qB\rho_{max}}} &=& \frac{q\rho_{max}}{e}\left(\frac{1}{\rho_{max}}-\frac{1}{q\rho_{max}}\right)\\
\left(\frac{1}{\rho_{max}}-\frac{1}{q\rho_{max}}\right)\frac{1}{B}&=& \W\left(\frac{q\rho_{max}}{e}\left(\frac{1}{\rho_{max}}-\frac{1}{q\rho_{max}}\right)\right)\\
B &=& -\frac{1-q}{q\rho_{max}\W\left(-\frac{1-q}{e}\right)}
\end{eqnarray}

To finish this section we summarize the model simplifications which we had to assume to gain the results:
\begin{enumerate}
\item Homogeneous population: all pedestrians have identical parameters. Microsimulations allow by construction to take account for a heterogeneous population by assigning each unit individual parameters. To allow analytical investigation we had to refrain from that. Obviously the values in our analysis would be the average or median value of any distribution. To test the theory simulations can be run with identical parameters for all pedestrians. 
\item One-dimensional movement: reducing the dimensionality from two to one spatial dimension introduces a fixed order between pedestrians which is required for the next simplification step of considering only nearest neighbors. %It is our experience that it works reasonably well to use the square root of a two-dimensional density where in equations derived above there is a one-dimensional density.
\item Only nearest neighbor interaction: the SFM originally assumed that each pedestrian in a simulation exerts a force on any other. The reduction to nearest neighbor interactions made the analytical investigation much easier, but obviously this step implies that results cannot be exact. It may be interesting in this regard that it was found that the approach to consider all other pedestrians as having an impact on a certain pedestrian does not lead to the most realistic speed-density relation, but that cutting forces at a certain number or at least suppressing them with more remote neighborhood degree is beneficial \cite{seyfried2006basics,Kretz2016inflection}. Thus the approximation of nearest neighbor interaction can also be interpreted as a model improvement. In computer implementations of simulation models usually the number of forces on a particular pedestrian is limited anyway to achieve reasonable computation times, which is another reason why this assumption is not only a simplification, but matches the pragmatic ways in which the SFM is actually applied.
\item Stationary state: it was assumed that movement is in a stationary state, i.e. macroscopic properties are stable over time. This is required by the aim of the work to relate {\em simple} macroscopic properties to model parameters. Without the system being stationary there are no simple properties. This assumption implies that one cannot expect results to describe situations exactly which include for example an acceleration phase.
\item Equal spacing between pedestrians instead of increasing distances toward rear end of queue: this is an assumption to which one can get arbitrarily close in a simulation by enlarging the size of the system.
\item By assuming that the queue is a {\em long} queue we implicitly assumed that the time delay from the first pedestrians in line having to accelerate from zero speed can be neglected. Thus our results cannot be transferred to queues consisting only a small number pedestrians.
\end{enumerate}

\section{Summary of Results and Discussion}
For $N$ pedestrians who stand still at the front end of a much longer single-file queue in front of a signal or service desk we find in the SFM with only nearest neighbor interaction for the length $L$ from the first to the $N$th pedestrian and the time $T$ it takes for discharge between the first and the $N$th pedestrian crossing a given cross section
\begin{eqnarray}
L &=& (N-1) B \ln(\alpha)                             \label{eq:L2} \\
T &=& - (N-1) \frac{B}{v_0} \W\left(-\frac{1}{\alpha e}\right)    \label{eq:T2}
\end{eqnarray}
with parameter combination $\alpha$ 
\begin{equation}
\alpha  = \frac{(1-\lambda) A \tau}{v_0} \label{eq:alpha1}
\end{equation}
where it ought not be forgotten that parameter $A$ is defined differently in different existing literature, compare equation (\ref{eq:AAtilde}).

In terms of maximum (stand still) density $\rho_{max}$ and capacity flow $j_c$ equations (\ref{eq:L2}) and (\ref{eq:T2}) can be rewritten
\begin{eqnarray}
\rho_{max} &=& \frac{1}{B \ln(\alpha)}                  \label{eq:rhomax2} \\
j_c &=& -\frac{v_0}{B} \frac{1}{\W(-\frac{1}{\alpha e})} \label{eq:jc2}
\end{eqnarray}

Defining a combination of observable parameters
\begin{equation}
q :=\frac{j_c}{v_0\rho_{max}} = \frac{L}{v_0T}\label{eq:q2} \\
\end{equation}
it is possible to rewrite equations (\ref{eq:rhomax2}) and (\ref{eq:jc2}) getting rid of either $\alpha$ or $B$:
\begin{eqnarray}
\frac{1}{q} &=& - \rho_{max} B W\!\left(-e^{-\left(1+\frac{1}{B\rho_{max}}\right)}\right)     \label{eq:noalpha2} \\
q &=& - \frac{\ln(\alpha)}{\W\left(-\frac{1}{\alpha e}\right)}           \label{eq:noB2}
\end{eqnarray}

These can be resolved for $\alpha$ and $B$, giving

\begin{eqnarray}
\alpha  &=& \left[-\W\left(-\frac{1-q}{e}\right) \frac{e}{1-q}\right]^{\frac{q}{1-q}} \label{eq:alpha2}\\
B &=& -\frac{1}{\W\left(-\frac{1-q}{e}\right)} \frac{1}{\rho_{max}} \frac{1-q}{q} \label{eq:B2}
\end{eqnarray}

Note that for $q$ it does not make a difference if $j_c$ refers to the total flow and $\rho_{max}$ to the line density or if $j_c$ refers to the specific flow and $\rho_{max}$ to the two dimensional density, since equation (\ref{eq:q2}) requires only an additional factor ``1=width/width'' to get from the former to the latter. Thus, in a certain sense parameter $q$ is independent of the dimension.

The first hope would now be that one could measure $v_0$, $\rho_{max}$ and $j_c$ in reality, from that compute a $q$ and with all these use equations (\ref{eq:alpha2}) and (\ref{eq:B2}) to compute parameters $\alpha$ and $B$. Then simulating with $\alpha$ and $B$ the same values as measured would result for $\rho_{max}$ and $j_c$ in the simulation. However, with all the approximations we have done chances for success are doubtful. We want to benefit from the results obtained from considering a 1d system also for 2d situations; and we do not seriously want to consider only the two nearest neighbors in a 2d simulation, but have at least 4, 6 or 8 pedestrians emitting a force on a certain pedestrian. Doing so would definitely change the results presented above for 1d as well as 2d situations, since the net value of the sum of forces will generally be larger. Thus, we cannot directly apply these equations. However, we can hope that the tendencies which they give are preserved even if we change fundamental properties of the system. In other words: in a calibration process the results derived above might allow an educated guess and tell in which direction parameters need to be changed so that the results change in a certain direction. 

Assume you already have done a simulation and you have obtained a certain $\rho_{max}$ and a certain $j_c$. You find that compared to your empirical data both -- $\rho_{max}$ as well as $j_c$ -- are too large. Then looking at equations (\ref{eq:rhomax2}) and (\ref{eq:jc2}) things are easy: increasing the value of $B$ will lower the value of both result attributes, and the same is true for parameter $\alpha$: larger $\alpha$ will lead to smaller $\rho_{max}$ and smaller $j_c$ as is shown in appendix \ref{app:1}.

The case when $\rho_{max}$ as well as $j_c$ both should result with larger values is just as trivial: lower the value(s) of $B$ or $\alpha$ or both. 

However, how should parameters be changed, if it is required to increase the value of $\rho_{max}$, but reduce $j_c$? Should the value of $\alpha$ be raised and the one of $B$ be lowered or the other way round? The answer is -- for the math see appendix \ref{app:2} -- that to achieve a higher value for $\rho_{max}$ and a lower value for $j_c$ the value of $\alpha$ needs to be reduced and the value of $B$ raised.

Summarized this suggests the following procedure for calibration of the SFM:
\begin{enumerate}
\item Measure or define reasonable values for desired speed $v_0$, capacity flow $j_c$ (or discharge rate $T$ for a given number $N$ of pedestrians), and maximum (stand still) density $\rho_{max}$ (or queue length $L$ for the same given number $N$ of pedestrians). There is no universally correct value for $L$ and $T$. Younger people may have different values than older ones. In emergency egress situations one may assume to find smaller values for $L$ as well as $T$ than if people are queuing and going to carry out some rather unpleasant duty. Make sure that $0<q<1$ in equation (\ref{eq:q2}).
\item With the help of equation (\ref{eq:B2}) compute a value for parameter $B$. The Lambert W function is implemented in numerical computing and computer algebra software like Matlab (``lambertw(-1, x)'') \cite{matlab1998mathworks}, Mathematica (``ProductLog[-1, x]'') \cite{wolfram1996mathematica} or Maple (``lambertw(-1, x)'') \cite{Maple}. One can also find plugins for Excel for download on the web. The easiest way to compute a few values is by using the (online) computational knowledge engine Alpha \cite{wolframalpha}. In any case make sure that the value from the lower real valued branch ($W_{-1}()$) is computed. 
\item With the help of equation (\ref{eq:alpha2}) compute a value for parameter combination $\alpha$. Make sure that $\alpha>1$. 
\item Choose values for parameters $\lambda$, $\tau$, and $A$, such that equation (\ref{eq:alpha1}) holds. Consider what is written above and in equation (\ref{eq:AAtilde}) about $A$ and $\tilde{A}$. Reasonable boundaries are $0\leq \lambda < 0.4$ (in \cite{johansson2007specification} and \cite{helbing2011pedestrian} values between 0.02 and 0.2 are given as resulting from empirical studies) and $0.05\leq \tau \approx<2.0$ s. Also consider that $4v_0 \tau / B \leq 1$ should hold or that at least the left side is only slightly larger than 1 since otherwise a pedestrian may visibly oscillate forth and back when approaching another \cite{kretz2015oscillations}.
\item Do simulations, evaluate in the simulation $\rho_{max}$ and $j_c$. And repeat
			\begin{itemize}
			\item if both, $\rho_{max}$ and $j_c$ need to be larger, reduce the value of $B$ and/or $\alpha$,
			\item if both, $\rho_{max}$ and $j_c$ need to be smaller, increase the value of $B$ and/or $\alpha$,
			\item if $\rho_{max}$ needs to be smaller and $j_c$ larger, reduce the value of $B$ and increase the value of $\alpha$,
			\item if $\rho_{max}$ needs to be larger and $j_c$ smaller, increase the value of $B$ and reduce the value of $\alpha$,
			\item simulate again and evaluate $\rho_{max}$ and $j_c$.
			\end{itemize}
\item Set the values of all other parameters in case you are working with a variant or an extension of the SFM. For this it is typically required to include more complex scenarios in the calibration procedure. For example the width of the distribution of desired speeds as well as the parameter that sets the impact strength of the relative velocity in elliptical specification II \cite{johansson2007specification} (called $VD$ or $\Delta t$) both are most relevant in bi- or multi-directional flow situations, so include scenarios with bi-directional and crossing flows. Remember that you are free to change the values of $\tau$, $\lambda$, and $A$ without spoiling the calibration, as long as the value of $\alpha$ in equation (\ref{eq:alpha1}) is conserved.
\end{enumerate}

\section{Verification and Example}
In this section we first check if the math presented above can be correct by trying it with a concrete example and simulating with the same model simplifications that were assumed. In a second step we release some of the simplifications, most importantly applying the calibration procedure in a 2d setting.

We configure a simulation of pedestrian dynamics \cite{Viswalk9} as close as possible to the model simplifications (a pedestrian ``sees'' forces only from two other pedestrians, very narrow distribution of desired speeds, circular specification of the SFM) and simulate a de facto one dimensional situation (50 cm wide corridor) where 1,000 pedestrians first walk toward a red traffic signal, stop there, and start moving when it turns green later. We measure the linear density of pedestrians as they wait in front of the red signal (number of pedestrians in a 100 m long section) and the discharge flow after the signal turns green (number of pedestrians passing within 100 seconds, measurement beginning 100 seconds after the signal turns green). We pretend we know nothing about the parameters of the SFM and use the empirical results of \cite{seyfried2005fundamental} ($v_0\approx 1.25$ m/s, $j_c\approx0.8$ s$^{-1}$, and $\rho_{max}\approx2.0$ m$^{-2}$) which give $q=0.32$ and from that with equations (\ref{eq:alpha2}) and (\ref{eq:B2}): $\alpha=2.7532$ and $B=0.4937$ m (in the software for this purpose it is $B_{soc, iso}=0.497$ m whereas $B_{soc, mean}$ is irrelevant since to disable the elliptical specification II it is required to set $A_{soc, mean}=0$). 

The first check is on the freedom to set the parameters $\tau$, $A$ and $\lambda$ that make up $\alpha$. The results as displayed in Table \ref{tab:1} show that in fact the flow (capacity) is constant that the maximum density varies only slightly for identical values of $\alpha$, such that the simulation in this aspect confirms the theory presented above. 

Looking at the particular values for flow and maximum density we see that the flow matches exactly expectations, by that increasing confidence in above's results further, but that the maximum density is slightly smaller than expected. This can be (at least) due to two causes: first, it can be a consequence of a violation of the (non-)oscillation condition $4v_0 \tau \leq B$ -- see \cite{kretz2015oscillations} -- i.e. pedestrians are not yet at rest at the time of measurement, but -- for some of the parameter sets clearly, for others barely visible -- still move forth and back at the time of measurement, by that requiring more space. Among others this is one of the reasons to not utilize the circular specification of the SFM alone, but to combine it with for example elliptic specification II which by construction suppresses oscillations. Second, our simulation system is not exactly one-dimensional, but it is a 2d simulation system artificially confined to something like a 1d system. This may have certain (small) effects which would be out of scope to be discussed here. In total we see the results to be in accordance with the theoretical considerations presented above.

\begin{table}[htbp]
\caption{Parameter choices and simulation results in a 1d setting. The pedestrians all had the same length of 0.456 m and the desired speed was sharply 1.25 m/s, and in all cases $B=0.4937$ m.}
\label{tab:1}
\centering
\begin{tabular}{ccc|cc}
\hline
$\tau$ [s] & $A_{Soc, Iso}$ [m/s$^2$] & $\lambda$ & $\rho_{max}$ [m$^{-1}$]& $j_c$ [s$^{-1}$] \\ \hline
0.4        & 3.390                    &  0.1      & 1.71                   & 0.8 \\
0.2        & 6.780                    &  0.1      & 1.91                   & 0.8 \\
0.15       & 9.041                    &  0.1      & 1.92                   & 0.8 \\
0.4        & 4.359                    &  0.3      & 1.82                   & 0.8 \\
\end{tabular}
\end{table}

The second test is in a 2d situation: 8,000 pedestrians walk down a 4 m wide strip. We aim at reproducing Weidmann's properties of pedestrian dynamics ($v_0=1.34$ m/s, $j_c=1.25$ (ms)$^{-1}$, and $\rho_{max}=5.4$ m$^{-2}$; implying $q=0.172$ and with equation (\ref{eq:alpha2}) $\alpha=1.44$). Compared to other literature data this capacity flow is small and maximum density moderate. We assign pedestrians a radius of 0.213 m $\pm$ 10\%, implying that 5.4 such (circle shaped) pedestrians cover an area of 0.77 m$^2$. This is near the maximum size (radius of 0.231 m) which allows to reach the mentioned maximum density without overlapping (densest packing of spheres). This raises the expectation that it might be difficult to reproduce the desired maximum density of 5.4 m$^{-2}$ with any set of parameters.

Going from 1d to 2d first bears the difficulty how to transfer equation (\ref{eq:B2}), since maximum density in 2d has a dimension of m${^-2}$ while $B$ needs the dimension m. We assumed -- what we believe to be the simplest assumption -- that the maximum density needs to be divided by a lane width $w_l$ and the explicit $\rho_{max}$ in equation (\ref{eq:B2}) needs to be replaced by $\rho_{max}/w_l$. As initial guess we assume a lane width of $w_l=0.5$ m. This leads to $B=1.02$ m.

Going from 1d to 2d it is furthermore anything but sure that the properties and relations derived above still hold. We want to test a) if as in 1d also in 2d capacity flow and maximum density only depend on $\alpha$, but not on how $A$, $\tau$, $\lambda$, and $v_0$ contribute to the value of $\alpha$, b) if equations (\ref{eq:rhomax2}) and (\ref{eq:jc2}) have a meaning in 2d, and c), if the calibration procedure as defined above can in fact be carried out in a 2d setting.

Another question in a 2d environment is, how many pedestrians should be taken into account as ``nearest neighbors'' for the SFM. Obviously two would be too few. Assuming a near to hexagonal (or triangular) walking formation \footnote{It is reported in \cite{porzycki2014pedestrian} that the sixth closest neighbor has the highest probability to walk exactly in front of a person. This can be seen as an indication in favor of this idea.} we choose six as number of influencing pedestrians.

The results as given in Table \ref{tab:2} support two conclusions: first, in 2d as in 1d $\rho_{max}$ and $j_c$ depend only on $\alpha$ and not on its factors. It was not clear at all that this fact would ``survive'' the expansion from one to two dimensions. Therefore this result is explicitly pleasant. Second, both $\rho_{max}$ as well as $j_c$ result much smaller than expected from the 1d theory. Obviously concerning this aspect the transgression from 1d to 2d brings a major change and with it a challenge for future work.

\begin{table}[htbp]
\caption{Parameter choices and simulation results in 2d setting}
\label{tab:2}
\centering
\begin{tabular}{ccc|cc}
\hline
$\tau$ [s] & $A_{Soc,Iso}$ [m/s$^2$] & $\lambda$ & $\rho_{max}$ [m$^{-2}$]& $j_c$ [(ms)$^{-1}$] \\ \hline
0.2        & 6.477                   &  0.1      & 1.40                  & 0.43 \\
0.4        & 3.239                   &  0.1      & 1.40                  & 0.43 \\
0.2        & 5.830                   &  0.0      & 1.41                  & 0.43 \\
\end{tabular}
\end{table}

The simplest way to bring the simulation results closer to the expected values is to assume larger values for $w_l$: $w_l=1.0$ m leads to $\rho_{max}=4.10$ m$^{-2}$ and $j_c=1.56$ (ms)$^{-1}$ such that one of the result parameters ($\rho_{max}$) is smaller than desired and the other one larger. This implies that the best parameters $A$ and $B$ may neither be both smaller nor both larger than the ones applied at this point ($A=3.9$ m/s$^2$ and $B=0.51$ m). Furthermore we apply the process and change parameters $A$ and $B$ following results of previous simulations as described above, keeping the values of $\tau$ and $\lambda$ constant. Table \ref{tab:3} gives an overview of those simulations where the resulting capacity flow is close to the desired value. It can easily be seen that indeed the value of parameter $A$ needs to be reduced and the value of parameter $B$ increased to get closer to the desired result. However, even for $B\rightarrow \infty$ the desired maximum density cannot be achieved for the given flow capacity. 

It may appear disappointing at first that the model parameters cannot be set such that the desired values of result parameters are met. However, at second thought this can also be interpreted as a positive property of the model: since the pedestrians in the simulation had a size (radius, to be precise) which implied a maximum density (densest packing of spheres) near the desired result for $\rho_{max}$ one may simply require from a model that it cannot be tweaked by parameter modifications to produce any result one could think of.

\begin{table}[htbp]
\caption{More parameter choices and simulation results in 2d setting. For all simulations it has been $\tau=0.2$ s and $\lambda=0.1$. }
\label{tab:3}
\centering
\begin{tabular}{cc|cc}
\hline
$A_{Soc,Iso}$ [m/s$^2$] & $B_{Soc,Iso}$ [m] & $\rho_{max}$ [m$^{-2}$]& $j_c$ [(ms)$^{-1}$] \\ \hline
30.0                    &  0.1              & 1.99                  & 1.19 \\
10.0                    &  0.3              & 2.82                  & 1.22 \\
5.0                     &  0.5              & 3.49                  & 1.20 \\
3.0                     &  1.0              & 4.02                  & 1.28 \\
2.5                     &  2.0              & 4.04                  & 1.25 \\
2.0                     &  10.0             & 4.43                  & 1.30 \\
1.9                     &  30.0             & 4.51                  & 1.33 \\
1.9                     &  50.0             & 4.49                  & 1.30 \\
1.9                     &  100.0            & 4.48                  & 1.29 \\
1.9                     &  10,000.0         & 4.46                  & 1.26 \\
\end{tabular}
\end{table}

Finally we checked once more that this depends only on $\alpha$ and not the parameters it is made of and varied the values of $\lambda$ and $\tau$ as in Table \ref{tab:2} adjusting $A_{soc,iso}$ such that $\alpha=0.2566$ is preserved. Again $j_c$ and $\rho_{max}$ remained nearly unchanged.

\section{Summary and Outlook}
In this contribution we were able to derive two formulas that assign values to a parameter ($B$) and a parameter combination ($\lambda$, $\tau$, $A$) for a simplified version of the SFM using as input only observable properties (stand still density, capacity flow, and free speed). Based on this we proposed a calibration procedure that can be applied requiring to know neither the derivation of the results as given in this work nor details of the SFM. We demonstrated the calibration procedure with a concrete example. As a result of the insight into fundamentals of the Social Force Model presented in this paper this calibration procedure is an {\em informed} procedure that exploits the model structure for calibration as opposed to optimization schemes which work irrespective of this. Nevertheless it may well be that a particular general optimization scheme can be combined with the particular information of the facing work to enhance the performance of an optimization (calibration) process.

Future work may concern mainly the transgression from one to two dimensions as well as the further calibration process. The approach presented in this work leaves freedom to choose some of the parameters. For extended variants of the SFM with additional parameters there is even more such freedom. To calibrate the values of these parameters empirical data from further walking situations (e.g. bi-directional, crossing) is required.

\section{Acknowledgments}
For the preparation of this contribution we used David Pritchard's Transportation Research Board template for LaTeX\cite{Pritchard2009Latex}.
%
% Non-BibTeX users please follow the syntax
% the syntax of "referenc.tex" for your own citations
\bibliographystyle{unsrt}
\bibliography{author}

\begin{thebibliography}{35}
\providecommand{\natexlab}[1]{#1}

\bibitem[{Helbing and Molnar(1995)}]{helbing1995social}
Helbing, D. and P.~Molnar, Social force model for pedestrian dynamics.
  \emph{Physical review E}, Vol.~51, No.~5, 1995, p. 4282.

\bibitem[{Helbing et~al.(2000)Helbing, Farkas, and
  Vicsek}]{Helbing2000simulating}
Helbing, D., I.~Farkas, and T.~Vicsek, {Simulating dynamical features of escape
  panic}. \emph{Nature}, Vol. 407, 2000, pp. 487--490.

\bibitem[{Johansson et~al.(2007)Johansson, Helbing, and
  Shukla}]{johansson2007specification}
Johansson, A., D.~Helbing, and P.~Shukla, Specification of the social force
  pedestrian model by evolutionary adjustment to video tracking data.
  \emph{Advances in Complex Systems}, Vol.~10, No. supp02, 2007, pp. 271--288.

\bibitem[{Helbing and Johansson(2011)}]{helbing2011pedestrian}
Helbing, D. and A.~Johansson, Pedestrian, Crowd and Evacuation Dynamics. In
  \emph{Encyclopedia of Complexity and Systems Science} (R.~Meyers, ed.),
  Springer New York, New York, NY, 2011, pp. 697--716.

\bibitem[{Wiedemann(1974)}]{wiedemann1974simulation}
Wiedemann, R., {Simulation des Strassenverkehrsflusses}. {Schriftenreihe des
  IfV}, Karlsruhe, 1974.

\bibitem[{Burstedde et~al.(2001)Burstedde, Klauck, Schadschneider, and
  Zittartz}]{burstedde2001simulation}
Burstedde, C., K.~Klauck, A.~Schadschneider, and J.~Zittartz, Simulation of
  pedestrian dynamics using a two-dimensional cellular automaton. \emph{Physica
  A: Statistical Mechanics and its Applications}, Vol. 295, No.~3, 2001, pp.
  507--525.

\bibitem[{Nishinari et~al.(2004)Nishinari, Kirchner, Namazi, and
  Schadschneider}]{nishinari2004extended}
Nishinari, K., A.~Kirchner, A.~Namazi, and A.~Schadschneider, Extended floor
  field CA model for evacuation dynamics. \emph{IEICE Transactions on
  information and systems}, Vol.~87, No.~3, 2004, pp. 726--732.

\bibitem[{Kretz et~al.(2008)Kretz, Kaufman, and
  Schreckenberg}]{kretz2008counterflow}
Kretz, T., M.~Kaufman, and M.~Schreckenberg, Counterflow extension for the
  FAST-model. In \emph{Cellular Automata}, Springer, 2008, pp. 555--558.

\bibitem[{Yu et~al.(2005)Yu, Chen, Dong, and Dai}]{yu2005centrifugal}
Yu, W., R.~Chen, L.~Dong, and S.~Dai, Centrifugal force model for pedestrian
  dynamics. \emph{Physical Review E}, Vol.~72, No.~2, 2005, p. 026112.

\bibitem[{Lakoba et~al.(2005)Lakoba, Kaup, and
  Finkelstein}]{lakoba2005modifications}
Lakoba, T., D.~Kaup, and N.~Finkelstein, Modifications of the
  Helbing-Molnar-Farkas-Vicsek social force model for pedestrian evolution.
  \emph{Simulation}, Vol.~81, No.~5, 2005, pp. 339--352.

\bibitem[{Yu and Johansson(2007)}]{yu2007modeling}
Yu, W. and A.~Johansson, Modeling crowd turbulence by many-particle
  simulations. \emph{Physical review E}, Vol.~76, No.~4, 2007, p. 046105.

\bibitem[{Parisi et~al.(2009)Parisi, Gilman, and
  Moldovan}]{parisi2009modification}
Parisi, D., M.~Gilman, and H.~Moldovan, A modification of the social force
  model can reproduce experimental data of pedestrian flows in normal
  conditions. \emph{Physica A: Statistical Mechanics and its Applications},
  Vol. 388, No.~17, 2009, pp. 3600--3608.

\bibitem[{Steffen(2010)}]{steffen2010modification}
Steffen, B., A modification of the social force model by foresight. In
  \emph{Pedestrian and Evacuation Dynamics 2008}, Springer, 2010, pp. 677--682.

\bibitem[{Zanlungo et~al.(2011)Zanlungo, Ikeda, and Kanda}]{zanlungo2011social}
Zanlungo, F., T.~Ikeda, and T.~Kanda, Social force model with explicit
  collision prediction. \emph{EPL (Europhysics Letters)}, Vol.~93, No.~6, 2011,
  p. 68005.

\bibitem[{Ratsamee et~al.(2012)Ratsamee, Mae, Ohara, Takubo, and
  Arai}]{ratsamee2012modified}
Ratsamee, P., Y.~Mae, K.~Ohara, T.~Takubo, and T.~Arai, Modified social force
  model with face pose for human collision avoidance. In \emph{Human-Robot
  Interaction (HRI), 2012 7th ACM/IEEE International Conference on}, IEEE,
  2012, pp. 215--216.

\bibitem[{Kretz et~al.(2016)Kretz, Lohmiller, and Schlaich}]{kretz2015social}
Kretz, T., J.~Lohmiller, and J.~Schlaich, The Social Force Model and its
  Relation to the Kladek Formula. In \emph{95th Annual Meeting of the
  Transportation Research Board}, 2016, \#16-1276.

\bibitem[{Chraibi et~al.(2010)Chraibi, Seyfried, and
  Schadschneider}]{chraibi2010generalized}
Chraibi, M., A.~Seyfried, and A.~Schadschneider, Generalized centrifugal-force
  model for pedestrian dynamics. \emph{Physical Review E}, Vol.~82, No.~4,
  2010, p. 046111.

\bibitem[{Campanella(2010)}]{campanella2010improving}
Campanella, M., Improving the Nomad microscopic walker model. In \emph{12th
  IFAC symposium on transportation systems, Redondo Beach, Sept. 2009}, IFAC,
  2010.

\bibitem[{Corless et~al.(1997)Corless, Jeffrey, and
  Knuth}]{corless1997sequence}
Corless, R., D.~Jeffrey, and D.~Knuth, A sequence of series for the Lambert W
  function. In \emph{Proceedings of the 1997 international symposium on
  Symbolic and algebraic computation}, ACM, 1997, pp. 197--204.

\bibitem[{Barry et~al.(2000)Barry, Parlange, Li, Prommer, Cunningham, and
  Stagnitti}]{barry2000analytical}
Barry, D., J.-Y. Parlange, L.~Li, H.~Prommer, C.~Cunningham, and F.~Stagnitti,
  Analytical approximations for real values of the Lambert W-function.
  \emph{Mathematics and Computers in Simulation}, Vol.~53, No.~1, 2000, pp.
  95--103.

\bibitem[{Chapeau-Blondeau and Monir(2002)}]{chapeau2002numerical}
Chapeau-Blondeau, F. and A.~Monir, Numerical evaluation of the Lambert W
  function and application to generation of generalized Gaussian noise with
  exponent 1/2. \emph{Signal Processing, IEEE Transactions on}, Vol.~50, No.~9,
  2002, pp. 2160--2165.

\bibitem[{Scott et~al.(2014)Scott, Fee, and Grotendorst}]{scott2014asymptotic}
Scott, T., G.~Fee, and J.~Grotendorst, Asymptotic series of generalized Lambert
  W function. \emph{ACM Communications in Computer Algebra}, Vol.~47, No. 3/4,
  2014, pp. 75--83.

\bibitem[{Weisstein(line)}]{WeissteinLambertW}
Weisstein, E., \emph{{Lambert W-Function}}, online, from MathWorld--A Wolfram
  Web Resource. \url{http://mathworld.wolfram.com/LambertW-Function.html}.

\bibitem[{Veberic(2010)}]{veberic2010having}
Veberic, D., Having fun with Lambert W (x) function. \emph{arXiv preprint
  arXiv:1003.1628}, 2010.

\bibitem[{Seyfried et~al.(2005)Seyfried, Steffen, Klingsch, and
  Boltes}]{seyfried2005fundamental}
Seyfried, A., B.~Steffen, W.~Klingsch, and M.~Boltes, The fundamental diagram
  of pedestrian movement revisited. \emph{Journal of Statistical Mechanics:
  Theory and Experiment}, Vol. 2005, No.~10, 2005, p. P10002.

\bibitem[{Seyfried et~al.(2006)Seyfried, Steffen, and
  Lippert}]{seyfried2006basics}
Seyfried, A., B.~Steffen, and T.~Lippert, Basics of modelling the pedestrian
  flow. \emph{Physica A: Statistical Mechanics and its Applications}, Vol. 368,
  No.~1, 2006, pp. 232--238.

\bibitem[{Kretz et~al.(2017)Kretz, Lohmiller, and
  Schlaich}]{Kretz2016inflection}
Kretz, T., J.~Lohmiller, and J.~Schlaich, The Inflection Point of the
  Speed-Density Relation and the Social Force Model. In \emph{Traffic and
  Granular Flow 2015} ({V. Knoop, W. Daamen, and S. Hoogendoorn}, ed.),
  2016-2017, p.~0, accepted for publication; preprint available at
  \url{http://arxiv.org/abs/1507.04935}.

\bibitem[{Mathworks(2014)}]{matlab1998mathworks}
Mathworks, \emph{MATLAB documentation}, 2014,
  \url{http://de.mathworks.com/help/matlab/}.

\bibitem[{Wolfram(1996)}]{wolfram1996mathematica}
Wolfram, S., \emph{The mathematica book}. Wolfram Media, Incorporated, 1996.

\bibitem[{Maplesoft(2014)}]{Maple}
Maplesoft, \emph{Maple User Manual - Maplesoft}, 2014,
  \url{www.maplesoft.com/view.aspx?sl=5883}.

\bibitem[{wol(2016)}]{wolframalpha}
\emph{Wolfram Alpha}, 2016, http://www.wolframalpha.com.

\bibitem[{Kretz(2015)}]{kretz2015oscillations}
Kretz, T., On oscillations in the Social Force Model. \emph{Physica A:
  Statistical Mechanics and its Applications}, Vol. 438, 2015, pp. 272--285.

\bibitem[{{PTV Group}(2016)}]{Viswalk9}
{PTV Group}, \emph{{PTV Vissim 9 manual}}, 2016, chapter 9, Viswalk.

\bibitem[{Porzycki et~al.(2014)Porzycki, Mycek, Luba{\'s}, and
  Was}]{porzycki2014pedestrian}
Porzycki, J., M.~Mycek, R.~Luba{\'s}, and J.~Was, Pedestrian Spatial
  Self-organization According to its Nearest Neighbor Position.
  \emph{Transportation Research Procedia}, Vol.~2, 2014, pp. 201--206.

\bibitem[{Pritchard(2009)}]{Pritchard2009Latex}
Pritchard, D., \emph{Transportation Research Board template for LaTeX}.
  \url{http://davidpritchard.org/archives/116}, 2009.

\end{thebibliography}
%%%%%%%%%%%%%%%%%%%%%%%%%%%%%%%%%%%%%%%%%%%%%%%%%%%%%%%%%%%%%%%%%%%%%%

%%%%%%%%%%%%%%%%%%%%%%%%%%%%%%%%%%%%%%%%%%%%%%%%%%%%%%%%%%%%%%%%%%%%%%
\pagebreak

\begin{appendix}
\section{Appendix}
\subsection{The function $-1/\W(-1/(e\alpha))$}\label{app:1}
\begin{figure}[htbp]
\includegraphics[width=6.5in]{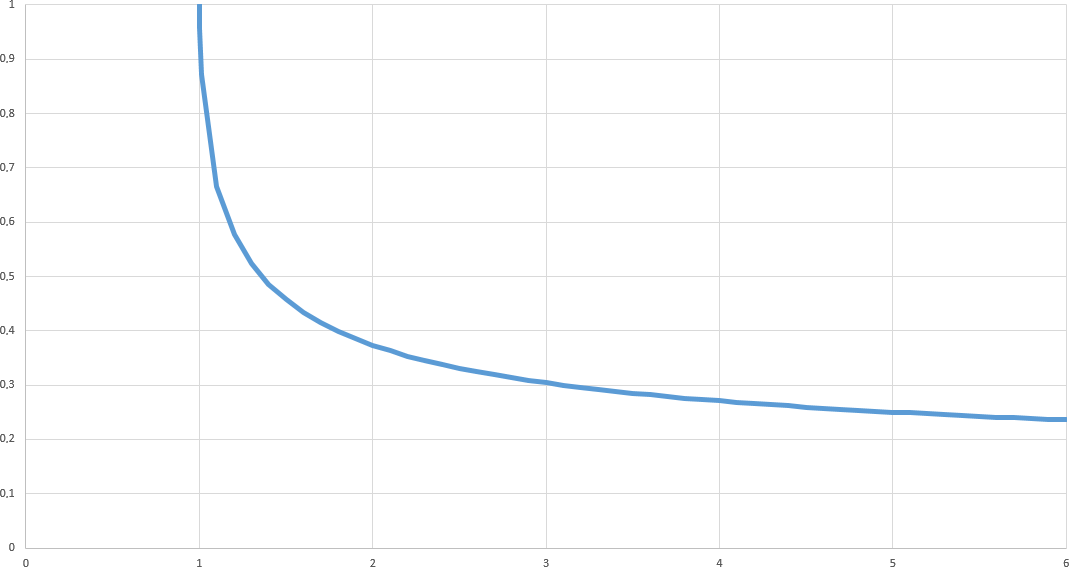}
\caption{$f(x)=-\frac{1}{\W\left(-\frac{1}{ex}\right)}$ with $x$ from 1 to 6.}
\label{fig:1overW}
\end{figure}

The easiest way to gain some insight into the properties of a function is to plot it, see Figure \ref{fig:1overW}. There the visual impression already suggests strongly that the first derivative by $\alpha$ is always negative. This can be proven easily
\begin{eqnarray}
f(y(\alpha)) &=& -\frac{1}{\W(y(\alpha))} \\
y(\alpha) &=& -\frac{1}{e\alpha}\\
\frac{df(y(\alpha))}{d\alpha} &=& \frac{1}{\W^2(y(\alpha))}\frac{d\W(y)}{dy}\frac{dy(\alpha)}{d\alpha}
\end{eqnarray}
Using 
\begin{equation}
\frac{d\W(y)}{dy} = \frac{\W(y)}{y(1+\W(y))} 
\end{equation}
see for example \cite{WeissteinLambertW}, we get
\begin{eqnarray}
\frac{df(y(\alpha))}{d\alpha} &=& \frac{1}{\W^2(y(\alpha))}\frac{\W(y(\alpha))}{y(\alpha)(1+\W(y(\alpha)))}\frac{1}{e\alpha^2}\\
&=& \frac{1}{\W(y(\alpha))}\left(-e\alpha\right)\frac{1}{(1+\W(y(\alpha)))}\frac{1}{e\alpha^2}\\
&=& \left(-\frac{1}{\alpha}\right)\left(\frac{1}{\W(y(\alpha))}\right)\left(\frac{1}{1+\W(y(\alpha))}\right)
\end{eqnarray}
Since all three top level brackets enclose negative values for all $\alpha>1$ the derivative is negative for all $\alpha>1$.

\subsection{To lower the value of $\rho_{max}$ and raise the value of $j_c$}\label{app:2}
To answer the question how $\alpha$ and $B$ need to be changed to achieve a smaller $\rho_{max}$ and a larger $j_c$ we imagine that we already have the right value for $\rho_{max}$ and want to raise the value of $j_c$ without changing $\rho_{max}$. So we have ``old'' parameters $\alpha$ and $B$ and ``new'' parameters $\alpha'$ and $B'$. Since $\rho_{max}$ should remain constant it is required that 
\begin{eqnarray}
\frac{1}{B'\ln(\alpha')}&=&\frac{1}{B\ln(\alpha)}\\
\alpha'&=&\alpha^{\frac{B}{B'}}
\end{eqnarray}
Note that if $B'>B$ then $\alpha'<\alpha$ and vice versa.
This allows to eliminate $\alpha'$ in the new $j_c'$:
\begin{eqnarray}
j_c'&=& -\frac{v_0}{B}\frac{1}{\W\left(-\frac{1}{e\alpha^{\frac{B}{B'}}}\right)}\\
f(\epsilon)&:=&\frac{j_c'}{j_c}=\frac{1}{1+\epsilon}\frac{\W\left(-\frac{1}{e\alpha}\right)}{\W\left(-\frac{1}{e\alpha^{\frac{1}{1+\epsilon}}}\right)} \text{   where }\\
1+\epsilon &:=&\frac{B'}{B}
\end{eqnarray}

Parameter $\epsilon$ quantifies the difference between old and new $B$; if $\epsilon<0$ then $B'<B$, otherwise $B'\geq B$. The hands-on method to see that $f(\epsilon)>1$ if $\epsilon<0$ and $f(\epsilon)<1$ if $\epsilon>0$ is to plot $f(\epsilon)$ for various $\alpha$ and thereby see that it appears to strictly monotonically decrease with growing $\epsilon$ and for all $\alpha$. A more mathematical approach is to compute the linear Taylor approximation of $f(\epsilon)$:
\begin{eqnarray}
\frac{df(\epsilon)}{d\epsilon}&=&-\left(\frac{1}{1+\epsilon}\right)^2\frac{\W\left(-\frac{1}{e\alpha}\right)}{\W\left(-\frac{1}{e\alpha^{\frac{1}{1+\epsilon}}}\right)}\left(1+\frac{1}{1+\epsilon}\frac{\ln(\alpha)}{1+\W\left(-\frac{1}{e\alpha^{\frac{1}{1+\epsilon}}}\right)}\right)\\
\frac{df}{d\epsilon}(0)&=&-\left(1+\frac{\ln(\alpha)}{1+\W\left(-\frac{1}{e\alpha}\right)}\right)
\end{eqnarray}

So we have
\begin{equation}
f(\epsilon)\approx 1 - \left(1+\frac{\ln(\alpha)}{1+\W\left(-\frac{1}{e\alpha}\right)}\right)\epsilon
\end{equation}
for small values of $\epsilon$, where the value inside the top level brackets is always positive for all $\alpha>1$, otherwise there would be some $a>1$ for which holds
\begin{eqnarray}
-a &=& \frac{\ln(\alpha)}{1+\W\left(-\frac{1}{e\alpha}\right)} \\
\W\left(-\frac{1}{e\alpha}\right) &=& -\frac{1}{a}\ln(\alpha)-1\\
-\frac{1}{e\alpha} &=&\left(-\frac{1}{a}\ln(\alpha)-1\right)e^{-\frac{1}{a}\ln(\alpha)-1}\\
\alpha^{\frac{1}{a}-1}&=&\frac{1}{a}\ln(\alpha)+1
\end{eqnarray}
which is not possible since in the last equation the left side is $<1$ and the right side $>1$ since $\alpha>1$.

Thus we could show that $j_c'<j_c$ if $B'>B$ (with a certain $\alpha'<\alpha$). Coming back to the original idea of an unmodified $\rho_{max}$ where if $B'>B$ it needs to be $\alpha'<\alpha$: choose some $\hat{B}$ with $B<\hat{B}<B'$ and it will obviously be
\begin{eqnarray}
\rho_{max}(\hat{B},\alpha') &>& \rho_{max}(B,\alpha) \\
       j_c(\hat{B},\alpha') &<&        j_c(B,\alpha) 
\end{eqnarray}
which shows that to have a chance to raise the value of $\rho_{max}$ and lower it for $j_c$ one needs to raise the value of $B$ and lower it for $\alpha$.
\end{appendix}
\end{document}